\documentstyle[pra,aps,amssymb,psfig,twocolumn]{revtex}
\begin{document}

\draft

\title{Quantum local-field corrections and spontaneous decay}
\author{Stefan Scheel, Ludwig Kn\"oll, Dirk-Gunnar Welsch}
\address{Theoretisch-Physikalisches Institut, 
Friedrich-Schiller-Universit\"at Jena, Max-Wien-Platz 1, 07743 Jena, Germany}
\author{Stephen M.~Barnett}
\address{Department of Physics and Applied Physics, University of
Strathclyde, Glasgow G4 0NG, Scotland}
\date{\today}
\maketitle

\begin{abstract}
A recently developed scheme [S. Scheel, L. Kn\"oll, and D.-G. Welsch,
Phys. Rev. A {\bf 58}, 700 (1998)] for quantizing the macroscopic
electromagnetic field in linear dispersive and absorbing dielectrics
satisfying the Kramers-Kronig relations is used to derive the quantum
local-field correction for the standard virtual-sphere-cavity model. 
The electric and magnetic local-field operators are shown to be consistent 
with QED only if the polarization noise is fully taken into
account. It is shown that the polarization fluctuations 
in the local field can dramatically change the spontaneous decay rate, 
compared with the familiar result obtained from the classical 
local-field correction. In particular, the spontaneous emission rate 
strongly depends on the radius of the
local-field virtual cavity.
\end{abstract}

\pacs{42.50.-p,42.50.Ct,42.50.Lc}
\narrowtext
\section{Introduction}
Spontaneous emission by an excited atom is one of the most studied
examples of a quantum process and may be attributed, at least in part,
to fluctuations in the electromagnetic vacuum \cite{milonni}. The
vacuum field is modified by the local environment and this, in turn,
leads to a modification of the spontaneous emission rate. In this way
the spontaneous emission rate can be changed by embedding the
radiating atom inside a dielectric host
\cite{glalew91,barhutlou92,barhutloumat96,juz97,tip97,knomuk89}, or by
changing the boundary conditions either by a cavity
\cite{hulhilkle85,jheandhinmesmoihar87,kosshi96,takuji97,tomlen97} or
a suitable surface \cite{yeugus96,barnes98}. Recent experiments have examined
the emission by atoms embedded in dielectric hosts
\cite{rikkes95,holland98,vrilag98} and have encouraged us to
re-examine the problem of local-field corrections to the bulk
modification of the spontaneous decay rate.

The total decay rate $\Gamma$ can be split into two parts,
\begin{equation} \label{0}
\Gamma = \Gamma^\perp + \Gamma^\| ,
\end{equation}
in which we associate the transverse decay rate $\Gamma^\perp$ 
and the longitudinal decay rate $\Gamma^\|$ with the
contributions of the transverse and longitudinal fields respectively.
The dielectric-induced modification of the 
spontaneous emission rate in free space
can be ascribed to two effects associated with the bulk (macroscopic)
field in the medium and the other arising from the local (microscopic)
field. The bulk field correction multiplies the 
rate by the refractive index at the transition frequency
\cite{barhutlou92,barhutloumat96}. Local-field corrections present
more of a problem and have a form that is strongly model-dependent.
For the Clausius-Mosotti model, which
introduces a virtual cavity surrounding the atom, a classical
treatment of the local-field corrections
leads to the form \cite{barhutlou92,barhutloumat96}
\begin{equation} \label{1}
\Gamma_{\rm cl}^\perp = 
\eta(\omega_A) 
\left| \frac{\epsilon(\omega_A)+2}{3} 
\right|^2
\Gamma_0
\end{equation}
for the transverse decay rate of an atom in a bulk dielectric
of refractive index  $n(\omega)$ $\!=\sqrt{\epsilon(\omega)}$
$\!=$ $\eta(\omega)$ $\!+$ $\!i\kappa(\omega)$.
In Eq.~(\ref{1}), $\Gamma_0$ $\!=$ 
$\!\omega_A^3 \mu^2/(3\pi c^3 \hbar \epsilon_0)$
is the free-space spontaneous emission rate, 
where $\omega_A$ and $\mu$ are, respectively, 
the atomic transition frequency and the dipole
transition matrix element. The local
field correction in Eq.~(\ref{1}) arises from writing the local
electric field in terms of the macroscopic electric field  
and the commonly used induced polarization field.
It does not, however, take account of the
fluctuating component of polarization associated with absorption
losses. In this paper we investigate the changes that arise within the
Clausius-Mosotti model when this fluctuating component is included.

Recently, a scheme for quantizing the electromagnetic field in an
arbitrary linear dielectric medium has been proven to be consistent
with QED \cite{schknowel98}. It relies on the introduction of an
appropriately chosen infinite set of basic-field operators
\cite{gruwel96,hoknowel98,matloubarjef95} and 
their connection to the electromagnetic field operators via the
classical Green function. This scheme is a generalization of the
approach introduced by Huttner and Barnett 
\cite{hutbar92} based on a Hopfield model
\cite{hop58} of a homogeneous dielectric using Fano diagonalization
\cite{fan56} to obtain collective (polariton) excitations of the
electromagnetic field, the polarization and the reservoir.
As we will show below, the scheme can also be applied to
local-field quantization that is consistent with QED
and used for calculating the modified spontaneous decay rate 
including the local quantum effects of the fluctuating medium 
polarization. 

The paper is organized as follows. After a short review of the
quantization scheme in Sect.~\ref{quantization} we introduce the
quantum local-field correction in Sect.~\ref{localfield}. We then
apply the scheme to the calculation of the spontaneous decay rates
in Sect.~\ref{emission} followed by some concluding remarks in
Sect.~\ref{conclusions}. Details of the calculation will be given in
the Appendix.


\section{Quantization scheme} \label{quantization}

We begin with a brief review of the quantization scheme used
throughout the paper. Further details can be found in
\cite{schknowel98,gruwel96,hoknowel98}. The spectral decomposition of
the electric and magnetic field operators is given by
\begin{eqnarray} \label{2}
\hat{\bf E}({\bf r})&=&\int_0^{\infty} d\omega 
\,\underline{\hat{\bf E}}({\bf r},\omega) + {\rm H.c.} ,
\\ \label{3}
\hat{\bf B}({\bf r})&=&\int_0^{\infty} d\omega 
\,\underline{\hat{\bf B}}({\bf r},\omega) +{\rm H.c.} ,
\end{eqnarray}
where $\underline{\hat{\bf E}}({\bf r},\omega)$ and
$\underline{\hat{\bf B}}({\bf r},\omega)$ satisfy Maxwell's equations
\begin{eqnarray} 
{\bf \nabla}\cdot \underline{\hat{\bf B}}({\bf r},\omega)&=&0 ,\\
{\bf \nabla}\cdot \left[ \epsilon_0 \epsilon({\bf r},\omega)
\underline{\hat{\bf E}}({\bf r},\omega) \right] &=&
\underline{\hat{\rho}}({\bf r},\omega) ,
\end{eqnarray}
\begin{eqnarray} \label{6}
{\bf \nabla} \times \underline{\hat{\bf E}}({\bf r},\omega) &=&
i\omega \underline{\hat{\bf B}}({\bf r},\omega) ,\\
{\bf \nabla} \times \underline{\hat{\bf B}}({\bf r},\omega) &=&
-i\frac{\omega}{c^2} \epsilon({\bf r},\omega) \underline{\hat{\bf E}}({\bf
r},\omega) + \mu_0 \underline{\hat{\bf j}}({\bf r},\omega) 
\end{eqnarray}
[$\epsilon({\bf r},\omega)$ $\!=$ $\!\epsilon_R({\bf r},\omega)$
$\!+$ $\!i\epsilon_I({\bf r},\omega)$ is the permittivity].
The operator noise current density $\underline{\hat{\bf j}}({\bf
r},\omega)$ and the operator noise charge density
$\underline{\hat{\rho}}({\bf r},\omega)$, which had to be introduced
in order to be consistent with the dissipation-fluctuation theorem,
are related to the noise polarization $\underline{\hat{\bf
P}}{^N}({\bf r},\omega)$ as
\begin{eqnarray} \label{8}
\underline{\hat{\bf j}}({\bf r},\omega)&=&-i\omega \underline{\hat{\bf
P}}{^N}({\bf r},\omega) ,\\
\underline{\hat{\rho}}({\bf r},\omega)&=& -{\bf \nabla} \cdot
\underline{\hat{\bf P}}{^N}({\bf r},\omega) 
\end{eqnarray}
and satisfy the equation of continuity
\begin{equation}
{\bf \nabla} \cdot \underline{\hat{\bf j}}({\bf r},\omega)=i\omega
\underline{\hat{\rho}}({\bf r},\omega) .
\end{equation}
The operator noise current density $\underline{\hat{\bf j}}({\bf r},\omega)$
is obtained from a bosonic vector field $\hat{\bf f}({\bf r},\omega)$,  
\begin{eqnarray}
\underline{\hat{\bf j}}({\bf r},\omega) &=& \omega
\sqrt{\frac{\hbar\epsilon_0}{\pi} \epsilon_I({\bf r},\omega)} \,
\hat{\bf f}({\bf r},\omega) ,\\ \label{12}
\left[ \hat{f}_i({\bf r},\omega), \hat{f}_j^\dagger({\bf r'},\omega')
\right] &=& \delta_{ij} \delta({\bf r-r'}) \delta(\omega-\omega') ,\\
\left[ \hat{f}_i({\bf r},\omega), \hat{f}_j({\bf r'},\omega')
\right] &=& \left[ \hat{f}_i^\dagger({\bf r},\omega),
\hat{f}_j^\dagger({\bf r'},\omega') \right] =0.
\end{eqnarray}
The quantization scheme implies that all electromagnetic field
operators can be expressed in terms of the basic fields 
$\hat{\bf f}({\bf r},\omega)$, which may be regarded as being the collective
excitations of the electromagnetic field, the medium polarization, and
the reservoir. For example, the electric-field operator $\underline{
\hat{\bf E}}({\bf r},\omega)$ satisfies the partial differential equation
\begin{equation}
{\bf \nabla} \times {\bf \nabla} \times \underline{\hat{\bf E}}({\bf
r},\omega) -\frac{\omega^2}{c^2} \epsilon({\bf r},\omega) 
\underline{\hat{\bf E}}({\bf r},\omega) 
=i\mu_0 \omega \underline{\hat{\bf j}}({\bf r},\omega),
\end{equation}
such that
\begin{equation} \label{15}
\underline{\hat{E}}_i({\bf r},\omega)=i\mu_0 \int d^3{\bf s} \,\omega
G_{ij}({\bf r},{\bf s},\omega) 
\underline{\hat{j}}_j({\bf s},\omega) ,
\end{equation}
where $G_{ij}({\bf r},{\bf s},\omega)$ is the tensor-valued Green
function of the classical problem. It can then be proven \cite{schknowel98}
that this quantization scheme is fully consistent with QED 
for arbitrary linear dielectrics.


\section{Quantum local-field correction} \label{localfield}

If we think of an atom located at some space point ${\bf r}_A$ inside
the dielectric, then the macroscopic field of Sect.~\ref{quantization}
will not, in fact, be the field felt by the atom. From classical
electrodynamics we know that we should introduce what is called the
local field at the location of the atom \cite{jackson}.
There are essentially two ways of introducing the local field. First, one
could cut out a {\it real} cavity \cite{glalew91} (most commonly 
a sphere) around the atom and calculate, in our scheme, 
the electric field inside the cavity according to Eq.~(\ref{15}). 
This would lead us to introduce the electric field operator
$\underline{\hat{\bf E}}{^{\rm loc}}({\bf r},\omega)$  by the relation  
\begin{equation}
\underline{\hat{E}}_i{^{\!\!\rm loc}}({\bf r},\omega)=i\mu_0 \int d^3{\bf s} \,
\omega G_{ij}^{\rm inh}({\bf r},{\bf s},\omega) \underline{\hat{j}}_j({\bf
s},\omega) ,
\end{equation}
where $G_{ij}^{\rm inh}({\bf r},{\bf s},\omega)$ is the Green function of 
the classical problem of an inhomogeneous medium that consists of
the real cavity surrounded by the dielectric in which the atom
is embedded.

To avoid the solution of the inhomogeneous problem, commonly a simpler
{\it virtual}-cavity model of Clausius-Mosotti-type is used. In this model
the local field is (approximately) related to the macroscopic field, 
which in our scheme can be obtained according Eq.~(\ref{15}), with the
Green function for the bulk-medium problem. 
In classical optics the electric field is usually
corrected by adding a term proportional to the polarization, 
\begin{equation} 
\label{17}
\underline{\bf E}^{\rm loc}({\bf r},\omega) = \underline{\bf E}({\bf
r},\omega) +\frac{1}{3\epsilon_0} \, \underline{\bf P}({\bf r},\omega) ,
\end{equation}
where
\begin{equation}
\label{18}
\underline{\bf P}({\bf r},\omega) = \epsilon_0 \left[
\epsilon({\bf r},\omega)-1\right] \underline{\bf E}({\bf r},\omega) ,
\end{equation}
from which it follows that
\begin{equation} \label{18a}
\underline{\bf E}^{\rm loc}({\bf r},\omega) 
= {\textstyle\frac{1}{3}}\left[
\epsilon({\bf r},\omega)+2 \right] \underline{\bf E}({\bf r},\omega) . 
\end{equation}
This classically corrected local field is just the field used
for the derivation of the rate formula (\ref{1}). 

In classical optics Eq.~(\ref{18}) corresponds to the
zero-temperature limit. It does not, however, 
hold in operator form in quantum optics. 
Regarding Eq.~(\ref{18}) as an operator equation would
strongly contradict the QED equal-time commutation relations
in general, so that the operator counterpart of Eq.~(\ref{18a}) will
not be a true electric field. In order to obtain a canonical operator 
equation, we have to complement Eq.~(\ref{18}) by a noise polarization
$\underline{\hat{\bf P}}{^N}({\bf r},\omega)$:
\begin{equation} \label{20}
\underline{\hat{\bf P}}({\bf r},\omega) =
\epsilon_0 [\epsilon({\bf r},\omega)-1] 
\underline{\hat{\bf E}}({\bf r},\omega) 
+\underline{\hat{\bf P}}{^N}({\bf r},\omega) .
\end{equation}
Using the quantization scheme in Sec.~\ref{quantization}, from 
Eq.~(\ref{8}) the noise
polarization $\underline{\hat{\bf P}}{^N}({\bf r},\omega)$ is given by 
\begin{equation} \label{21}
\underline{\hat{\bf P}}{^N}({\bf r},\omega)= - \frac{1}{i\omega} \,
\underline{\hat{\bf j}}({\bf r},\omega) = i
\sqrt{\frac{\hbar\epsilon_0}{\pi} \epsilon_I({\bf r},\omega)} \,
\hat{\bf f}({\bf r},\omega) .
\end{equation}
Combining Eqs.~(\ref{17}) and (\ref{20}), a  
local electric-field operator 
\begin{eqnarray}
\label{22}
\underline{\hat{\bf E}}{^{\rm loc}}({\bf r},\omega)
= {\textstyle\frac{1}{3}}\left[\epsilon({\bf r},\omega)+2\right]
\underline{\hat{\bf E}}({\bf r},\omega) 
+ \frac{1}{3\epsilon_0} \,\underline{\hat{\bf P}}{^N}({\bf r},\omega)
\end{eqnarray}
can be introduced, with $\underline{\hat{\bf P}}{^N}({\bf r},\omega)$
being given in Eq.~(\ref{21}).

In order to prove the consistency of the quantized local field with
QED, we compute the (equal-time) commutation  
relation between the fundamental local fields 
$\hat{\bf E}{^{\rm loc}}({\bf r})$ and
$\hat{\bf B}{^{\rm loc}}({\bf r})$. For this purpose we note 
that the local magnetic-field operator is  
given by, on using Eqs.~(\ref{6}),
\begin{eqnarray}
\label{24}
\underline{\hat{\bf B}}{^{\rm loc}}({\bf r},\omega)
= {\bf \nabla} \times
{\cal P} \frac{1}{i\omega}
\,\underline{\hat{\bf E}}{^{\rm loc}}({\bf r},\omega),
\end{eqnarray}
where the symbol ${\cal P}$ stands for the principal part. 
Recalling Eqs.~(\ref{2}) and (\ref{3}),
the local-field operators in real space are
\begin{eqnarray}
\label{25}
\hat{\bf E}{^{\rm loc}}({\bf r}) =
\int_0^{\infty} d\omega
\,\underline{\hat{\bf E}}{^{\rm loc}}({\bf r},\omega)
+ {\rm H.c.}
\end{eqnarray}
and
\begin{eqnarray} 
\label{26}
\hat{\bf B}{^{\rm loc}}({\bf r}) = 
\int_0^{\infty} d\omega 
\,\underline{\hat{\bf B}}{^{\rm loc}}({\bf r},\omega)
+ {\rm H.c.}. 
\end{eqnarray}
Expressing the local electric and magnetic fields in terms
of the basic fields $\hat{\bf f}({\bf r},\omega)$,
from the calculation given in App.~\ref{appendix} it is found that 
\begin{eqnarray}
\label{27}
\lefteqn{
\left[ \hat{E}_i{^{\!\!\rm loc}}({\bf r}), \hat{B}_k{^{\!\!\rm loc}}({\bf r'})
\right] 
}
\nonumber\\&&\hspace{2ex}
= \, - \frac{i\hbar}{\epsilon_0} \epsilon_{ikl} \partial_l^{r}
\delta({\bf r}-{\bf r'})
\left\{1+{\textstyle\frac{1}{9}}\big[\epsilon({\bf r},0)-1\big]\right\} ,
\end{eqnarray}
and it is easily seen that
\begin{equation}
\left[ \hat{E}_i{^{\!\!\rm loc}}({\bf r}), 
\hat{E}_k{^{\!\!\rm loc}}({\bf r'})\right] 
= \left[ \hat{B}_i{^{\!\!\rm loc}}({\bf r}),
\hat{B}_k{^{\!\!\rm loc}}({\bf r'})\right] = 0 .
\end{equation}
The result reveals that the local-electric field operator (\ref{22})
and the associated magnetic field operator (\ref{24}) can 
be regarded as being consistent with quantum theory,
provided that the (real) static permittivity 
$\epsilon_S({\bf r})$ $\!=$ $\!\epsilon({\bf r},0)$
satisfies the condition  
\begin{eqnarray}
\label{27a}
\frac{\epsilon_S({\bf r})}{10}\ll 1.
\end{eqnarray}
Equivalently, the static refractive index $n_S({\bf r})$ $\!=$
$\!\sqrt{\epsilon_S({\bf r})}$ must be small compared with 
$\sqrt{10}$ $\!\approx$ $\!3.16$.

It should be noted that a term proportional to the $\delta$ function
$\delta(\omega)$ can be added to the right-hand side of 
Eq.~(\ref{24}) in order to recover Ampere's law when
the equation is multiplied by $\omega$.  
Obviously, this ambiguity reflects the fact that the 
static magnetic field cannot be inferred from the 
static electric field. From a simple calculation it can be 
shown that such a term does not change the commutation relation 
(\ref{27}). Since it is only relevant at zero frequency, it
does not play any role in the calculation of the decay at transition
frequency $\omega_A$.  

In order to take into account a possible deviation of the 
symmetry of the material from cubic symmetry, a structure constant
$s$ can be included in Eq.~(\ref{17}) such that \cite{bowdow93}
\begin{equation}
\underline{\bf E}^{\rm loc}({\bf r},\omega) = \underline{\bf E}({\bf
r},\omega) +\frac{1}{\epsilon_0}
\left[ \frac{1}{3} +s\right] \, \underline{\bf P}({\bf r},\omega).
\end{equation}
Regarding this equation as an operator equation with 
$\hat{\bf P}({\bf r},\omega)$ from Eq.~(\ref{20})
and following the line in Appendix~\ref{appendix}, it can be seen that 
Eq.~(\ref{27}) changes to 
\begin{eqnarray}
\lefteqn{
\left[ \hat{E}_i{^{\!\!\rm loc}}({\bf r}), \hat{B}_k{^{\!\!\rm loc}}({\bf r'})
\right] 
}
\nonumber\\&&\hspace{2ex}
= \, - \frac{i\hbar}{\epsilon_0} \epsilon_{ikl} \partial_l^{r}
\delta({\bf r}-{\bf r'})
\left\{1+\frac{\alpha^2}{9}\big[\epsilon({\bf
r},0)-1\big]\right\} ,
\end{eqnarray}
where the parameter $\alpha$ in related to $s$ by
\begin{equation}
\alpha= 1+3s .
\end{equation}
Thus, consistency with quantum theory is achieved, if the condition
\begin{equation}
\epsilon_S({\bf r}) \ll 9\alpha^{-2}+1
\end{equation}
is fulfilled.


\section{Spontaneous decay rate} \label{emission}

The spontaneous decay rate of a (two-level) atom at transition frequency 
$\omega_A$ placed at point ${\bf r}_A$ is given by
\begin{equation}
\label{28}
\Gamma=\frac{2\pi}{\hbar^2} \int d\omega \, \mu_i \langle 0 |
\underline{\hat{E}}_i{^{\!\!\rm loc}}({\bf r},\omega)
\underline{\hat{E}}_j{^{\!\!\rm loc\dagger}}({\bf r}_A,\omega_A) | 0 \rangle 
\mu_j 
\end{equation}
(${\bf r}$ $\!\to$ $\!{\bf r}_A$).
In what follows we consider a homogeneous bulk
material, i.e., $\epsilon({\bf r},\omega)$ $\!\equiv$
$\!\epsilon(\omega)$,
and assume that the inequality (\ref{27a}) is fulfilled. 
Using Eq.~(\ref{22}), the vacuum expectation value of the local
electric-field operators in the limit 
\mbox{${\bf r}$ $\!\to$ $\!{\bf r}_A$}
can be written as the sum
\begin{eqnarray} \label{31} 
\lefteqn{
\langle 0 | \underline{\hat{E}}_i{^{\!\!\rm loc}}({\bf r},\omega)
\underline{\hat{E}}_j{^{\!\!\rm loc\dagger}}({\bf r}_A,\omega') | 0 \rangle
}
\nonumber \\ && \hspace{2ex} 
=\,\frac{\epsilon(\omega)\!+\!2}{3}\frac{\epsilon^\ast(\omega')\!+\!2}{3}
\langle 0 |
\underline{\hat{E}}_i({\bf r},\omega)
\underline{\hat{E}}_j{^{\!\!\dagger}}({\bf r}_A,\omega') | 0 \rangle 
\nonumber \\ && \hspace{4ex}
+\,\frac{1}{9\epsilon_0^2} \, \langle 0 |
\underline{\hat{P}}_i{^{\!\!N}}({\bf r},\omega) 
\underline{\hat{P}}_j{^{\!\!N\dagger}}({\bf r}_A,\omega') | 0 \rangle
\nonumber \\ && \hspace{4ex} 
+\,\frac{\epsilon(\omega)\!+\!2}{9\epsilon_0}\, \langle 0 |
\underline{\hat{E}}_i({\bf r},\omega)
\underline{\hat{P}}_j{^{\!\!N\dagger}}({\bf r}_A,\omega') | 0 
\rangle 
\nonumber \\ && \hspace{4ex}
+\,\frac{\epsilon^\ast(\omega')\!+\!2}{9\epsilon_0}
\,\langle 0 | \underline{\hat{P}}_i{^{\!\!N}}({\bf r},\omega)
\underline{\hat{E}}_j{^{\!\!\dagger}}({\bf r}_A,\omega') | 0 \rangle ,
\end{eqnarray}
where $\underline{\hat{\bf E}}({\bf r},\omega)$ and
$\underline{\hat{\bf P}}{^N}({\bf r},\omega)$
are given by Eqs.~(\ref{15}) and (\ref{21}) respectively.
The Green function $G_{ij}({\bf r},{\bf r}_A,\omega_A)$ for 
the bulk material in the limit ${\bf r}$ $\!\to$ ${\bf r}_A$
has the form 
\begin{eqnarray} \label{36}
G_{ij}({\bf R},\omega_A)
= G^\perp_{ij}({\bf R},\omega_A) + G^\|_{ij}({\bf R},\omega_A)
\end{eqnarray}
(${\bf R}$ $\!=$ $\!{\bf r}$ $\!-$ $\!{\bf r}_A$), where
\begin{eqnarray} \label{36a} 
\lefteqn{
G^\perp_{ij}({\bf R},\omega_A)
= \frac{1}{4\pi} \left\{ \frac{R_i R_j}{2R^3} +\frac{\delta_{ij}}{2R}
\right.
}
\nonumber \\ && \hspace{4ex}
\left.
+ \, \frac{2i\omega_A}{3c} \big[ \eta(\omega_A) +i\kappa(\omega_A) \big]
\delta_{ij} \right\} +{\cal O}(R)
\end{eqnarray}
and
\begin{eqnarray} \label{36b} 
\lefteqn{
G^\|_{ij}({\bf R},\omega_A)
= - \frac{c^2}{4\pi\omega_A^2 \epsilon(\omega_A)}
}
\nonumber \\ && \hspace{4ex}\times
\left[ \frac{4\pi}{3} \delta_{ij} \delta({\bf R})
+ \left( \delta_{ij} -\frac{3R_i R_j}{R^2}
\right) \frac{1}{R^3} \right] +{\cal O}(R)
\end{eqnarray}
are the transverse and longitudinal parts respectively.
We see that the real part of the transverse Green function as well as the
longitudinal part itself 
diverge as \mbox{$R$ $\!\to$ $\!0$}, 
reflecting the fact that a macroscopic approach is 
valid only to some appropriately fixed scale $\bar{R}$ which
exceeds the average distance of two atoms in the dielectric. 
Following \cite{barhutloumat96}, we average the divergent
terms in the decay rate over a sphere of radius $\bar{R}$.
We will see later that one can estimate an upper 
bound for $\bar{R}$ from the decay rate itself. 

The first term on the right-hand side in Eq.~(\ref{31}) gives the 
contribution to the decay rate with the classically corrected local field
\cite{barhutlou92,barhutloumat96},
\begin{eqnarray}
\lefteqn{
\Gamma_{\rm cl} = \Gamma^\perp_{\rm cl} + \Gamma^\|_{\rm cl}
}
\nonumber\\&&\hspace{1ex}
= \Gamma_0 \left| \frac{\epsilon(\omega_A)\!+\!2}{3} \right|^2
\Bigg[ \eta(\omega_A)
+\frac{3\epsilon_I(\omega_A)}{2|\epsilon(\omega_A)|^2} \left(
\frac{c}{\omega_A \bar{R}} \right)^3 \Bigg],
\end{eqnarray}
with the transverse rate $\Gamma^{\perp}_{\rm cl}$
being given in Eq.~(\ref{1}).
The second term in Eq.~(\ref{31}) is purely a contribution of the
noise polarization field and is given by  
\begin{eqnarray} 
\lefteqn{
\frac{1}{9\epsilon_0^2} \, \langle 0 |
\underline{\hat{P}}_i{^{\!\!N}}({\bf r},\omega) 
\underline{\hat{P}}_j{^{\!\!N\dagger}}({\bf r}_A,\omega') | 0 \rangle
}
\nonumber \\ && \hspace{4ex}
=\,\frac{\hbar \epsilon_I(\omega)}{9\pi\epsilon_0} \,
\delta_{ij} \delta({\bf R}) \,\delta(\omega-\omega') .
\end{eqnarray}
The cross-terms mixing the macroscopic electric field and the 
noise polarization field give rise to the contribution 
\begin{eqnarray} 
\lefteqn{
\frac{\epsilon(\omega)+2}{9\epsilon_0} \langle 0 |
\underline{\hat{E}}_i({\bf r},\omega)
\underline{\hat{P}}_j{^{\!\!N\dagger}}({\bf r}_A,\omega') | 0 \rangle
}
\nonumber \\ && \hspace{2ex}
+ \, \frac{\epsilon^\ast(\omega')+2}{9\epsilon_0}
\langle 0 | \underline{\hat{P}}_i{^{\!\!N}}({\bf r},\omega)
\underline{\hat{E}}_j{^{\!\!\dagger}}({\bf r}_A,\omega') | 0 \rangle
\nonumber \\ && 
=\,\frac{2\omega^2 \hbar}{3\pi c^2 \epsilon_0} 
\epsilon_I(\omega) {\rm Re}\!\left[ \frac{
\epsilon(\omega)\!+\!2}{3} G_{ij}({\bf R},\omega) \right]
\delta(\omega\!-\!\omega') .
\end{eqnarray}
Hence, the total decay rate reads
\begin{eqnarray} \label{35}
\lefteqn{
\Gamma = \Gamma_{\rm cl} 
+\frac{2}{9\hbar\epsilon_0} \epsilon_I(\omega_A) \mu_i \mu_j \,
\overline{\delta_{ij}\delta({\bf R})} 
}
\nonumber \\ &&\hspace{2ex}
+\,\frac{4}{3\hbar\epsilon_0} \frac{\omega_A^2}{c^2}
\epsilon_I(\omega_A) \mu_i \mu_j {\rm Re}\!\left[ \frac{
\epsilon(\omega_A)\!+\!2}{3} \,\overline{G_{ij}({\bf R},\omega_A)}
\right].
\end{eqnarray}

Equation (\ref{35}) is remarkable in the sense that 
inclusion of the polarization noise in the local field 
gives rise to a term that only results from that noise
and leads to a dependence of the decay rate on the real part 
of the Green function. We average the $\delta$ tensor
\begin{eqnarray} \label{35a}
\delta_{ij}\delta({\bf R})
= \delta^\perp_{ij}({\bf R}) + \delta^\|_{ij}({\bf R})
\end{eqnarray}
and the Green tensor (\ref{36}) over the sphere and obtain 
\begin{eqnarray} \label{35b}
\overline{\delta_{ij}^\perp({\bf R})}
= 2 \overline{\delta_{ij}^\|({\bf R})}
= \frac{1}{2\pi\bar{R}^3} \, \delta_{ij},
\end{eqnarray}
\begin{eqnarray}
{\rm Re}\, \overline{G^\perp_{ij}({\bf R},\omega_A)}
&=& \left[\frac{1}{4\pi\bar{R}}
-\frac{\omega_A \kappa(\omega_A)}{6\pi c} 
\right] \delta_{ij},
\\
{\rm Re}\, \overline{G^\|_{ij}({\bf R},\omega_A)}
&=& -\frac{c^2\epsilon_R(\omega_A)}
{4\pi\omega_A^2 |\epsilon(\omega_A)|^2 \bar{R}^3}
\,\delta_{ij}, 
\\
{\rm Im}\, \overline{G^\perp_{ij}({\bf R},\omega_A)}
&=& \frac{\omega_A \eta(\omega_A)}{6\pi c} \,\delta_{ij},
\\
{\rm Im}\, \overline{G^\|_{ij}({\bf R},\omega_A)}
&=& \frac{c^2\epsilon_I(\omega_A)}{4\pi \omega_A^2 |\epsilon(\omega_A)|^2
\bar{R}^3} \, \delta_{ij}\,,
\end{eqnarray}
and Eq.~(\ref{35}) can be given in the form of Eq.~(\ref{0}), 
where $\Gamma^\perp$ and $\Gamma^\|$ read as
\begin{eqnarray} \label{39}
\lefteqn{
\Gamma^\perp = \Gamma_0 \Bigg\{ 
\eta(\omega_A) \bigg[\left|
\frac{\epsilon(\omega_A)\!+\!2}{3} \right|^2
- \frac{2\epsilon^2_I(\omega_A)}{9} \bigg]
}
\nonumber \\ && \hspace{5ex}  
 -\,\epsilon_I(\omega_A) \left[ \epsilon_R(\omega_A)+2 \right]
\bigg[ \frac{2}{9}\,\kappa(\omega_A)-\frac{c}{3\omega_A \bar{R}} \bigg]
\nonumber \\ && \hspace{5ex}
+ \, \frac{c^3 \epsilon_I(\omega_A)}{3 \omega_A^3 \bar{R}^3}
\Bigg\} 
\end{eqnarray}
and
\begin{eqnarray} \label{40}
\lefteqn{
\Gamma^\| = 
\frac{\mu^2 \epsilon_I(\omega_A)}{2\pi\hbar\epsilon_0
\bar{R}^3 |\epsilon(\omega_A)|^2} \Bigg\{ \left|
\frac{\epsilon(\omega_A)\!+\!2}{3} \right|^2
+\frac{|\epsilon(\omega_A)|^2}{9} 
}
\nonumber \\ && \hspace{6ex}
-\,\frac{2}{9} \,\epsilon_R(\omega_A) 
\left[\epsilon_R(\omega_A)\!+\!2\right] -\frac{2}{9} 
\,\big[\epsilon_I(\omega_A) \big]^2  \Bigg\} .
\end{eqnarray}

\begin{figure}[h]
\psfig{file=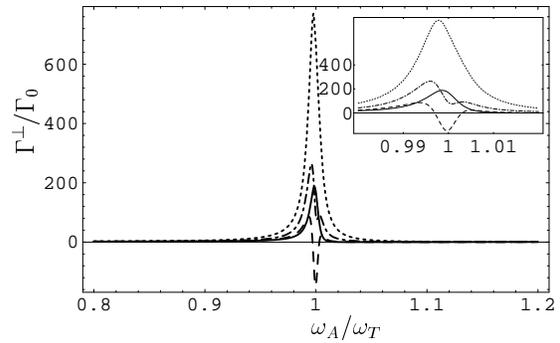,width=8cm}
\caption{\label{gamma001} 
The transverse decay rate $\Gamma^\perp$ is shown as a function
of the transition frequency $\omega_A$ for $\gamma$ $\!=$ $\!0.01$
and $r$ $\!=$ $\!10$ (dashed curve), $r$ $\!=$ $\!20$ (broken curve)
and $r$ $\!=$ $\!30$ (dotted curve). 
For comparison, the rate without quantum local-field correction
\protect\cite{barhutlou92} is shown (solid curve).
Since for \protect$r$ $\!=$ $\!10$ (broken curve) $\Gamma^\perp$
becomes negative, this case must be excluded from consideration
(cf. Fig.~\protect\ref{rmax}).}
\end{figure}

\begin{figure}[h]
\psfig{file=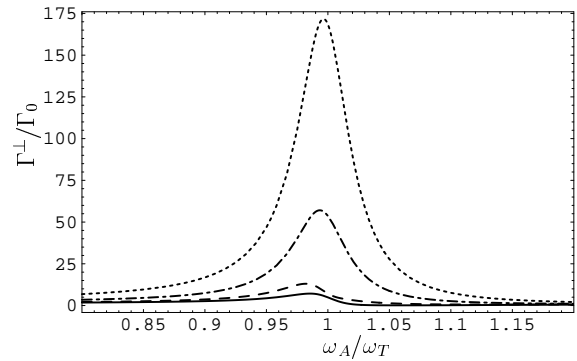,width=8cm}
\caption{\label{gamma005} 
The transverse decay rate $\Gamma^\perp$ is shown as a function
of the transition frequency $\omega_A$ for $\gamma$ $\!=$ $\!0.05$
and $r$ $\!=$ $\!10$ (dashed curve), $r$ $\!=$ $\!20$ (broken curve),
and $r$ $\!=$ $\!30$ (dotted curve).
For comparison, the rate without quantum local-field correction
\protect\cite{barhutlou92} is shown (solid curve).
}
\end{figure}

\begin{figure}[h]
\psfig{file=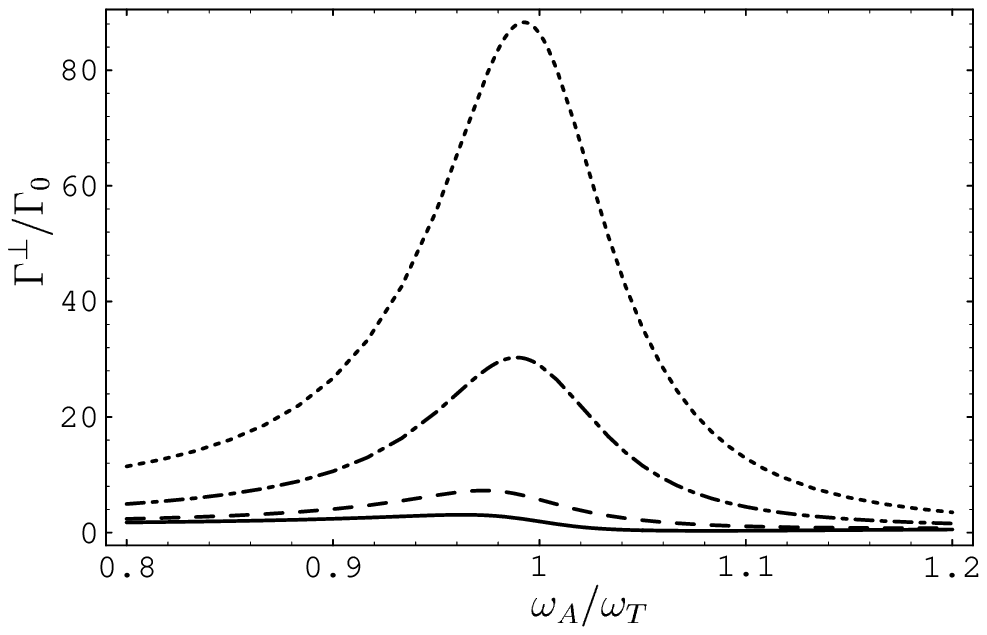,width=8cm}
\caption{\label{gamma01} 
The transverse decay rate $\Gamma^\perp$ is shown as a function
of the transition frequency $\omega_A$ for $\gamma$ $\!=$ $\!0.1$
and $r$ $\!=$ $\!10$ (dashed curve), $r$ $\!=$ $\!20$ (broken curve),
and $r$ $\!=$ $\!30$ (dotted curve).
For comparison, the rate without quantum local-field correction
\protect\cite{barhutlou92} is shown (solid curve).
} 
\end{figure}

The modifications near a medium resonance are clear. 
Note that owing to the quantum local-field correction the unspecified 
parameter $\bar{R}$ also enters into the transverse decay rate.
In order to compare our canonical result with that obtained using the
classically corrected local field, we use, for comparison, the same
Lorentz model for the permittivity of a single-resonance medium
as in \cite{barhutlou92,barhutloumat96}, 
\begin{equation}
\epsilon(\omega)=1+\frac{(0.46\,\omega_T)^2}{\omega_T^2-\omega^2
-i\gamma\omega} \,,
\end{equation}
where $\omega_T$ is the resonance frequency of the medium.
Figures \ref{gamma001}--\ref{gamma01} show the transverse 
decay rate $\Gamma^\perp$ with and without quantum local-field corrections  
as a function of the atomic transition frequency $\omega_A$
for different values of the damping parameter of the medium, $\gamma$, 
and the parameter
\begin{equation} \label{41}
r =\frac{\lambda_T}{\bar{R}} \,.
\end{equation}
First of all, for small $r$, i.e., large virtual-cavity radius $\bar{R}$ 
one observes little reduction of spontaneous decay for frequencies
$\omega_A$ just above the resonance frequency $\omega_T$. Its possible
applications in semiconductor physics and solid-state physics has
already been discussed \cite{yab87}. 

The greatest difference between the quantum-me\-cha\-nically and classically
corrected transverse decay rates $\Gamma^\perp$ and $\Gamma^\perp_{\rm cl}$,  
respectively, arises near the medium resonance when $\gamma$ is small. 
Both the imaginary part of the permittivity and the real part can take very
large values for $\omega_A$ $\!\approx$ $\!\omega_T$
and in consequence $\Gamma^\perp$ can drastically change compared with 
$\Gamma^\perp_{\rm cl}$. Obviously, in the resonance regime 
the noise polarization essentially contributes to the local-field
and therefore strongly influences $\Gamma^\perp$.  
For small values of $\gamma$ both qualitative and
quantitative differences between the rates $\Gamma^\perp$ and
$\Gamma^\perp_{\rm cl}$ are observed (Fig.~\ref{gamma001}). 
With increasing value of $\gamma$ the two rates become 
less different from each other, the 
changes being quantitative rather than qualitatitive
(compare Fig.~\ref{gamma001} with Fig.~\ref{gamma01}). 

\begin{figure}[h]
\psfig{file=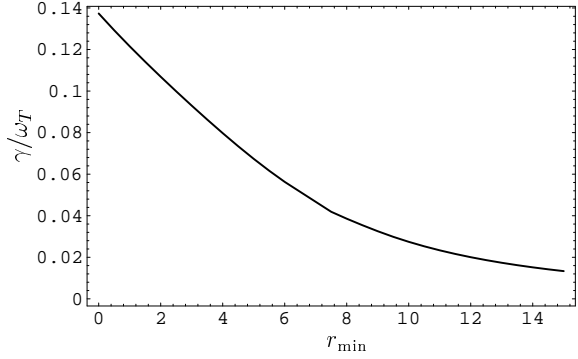,width=8cm}
\caption{\label{rmax} 
The lower bound $r_{\rm min}$ of the parameter $r$, Eq.~(\protect\ref{41}),
is shown as a function of the damping parameter $\gamma$.
The region below the curve is the part where the transverse decay rate
$\Gamma^\perp$ may take negative values and is therefore forbidden.} 
\end{figure}

Whereas $\Gamma^\perp_{\rm cl}$ is 
independent of the radius $\bar{R}$ of the virtual cavity, 
$\Gamma^\perp$ sensitively depends on $\bar{R}$. To fix the value of
$\bar{R}$ that is undetermined in the Clausius-Mosotti model,
experimental data could be used in principle 
(for recent experiments on spontaneous emission, see, e.g.,
\cite{rikkes95,holland98,vrilag98}). 
It is worth noting that our
theory gives an upper bound $\bar{R}_{\rm max}$, i.e.,
a lower bound $r_{\rm min}$ for the parameter $r$,
because of the fact that $\Gamma^\perp$ cannot be negative.
As already mentioned, the limit ${\bf r}$
$\!\to$ $\!{\bf r}_A$ in Eq.~(\ref{28}) 
cannot be performed and averaging over a sphere of radius $\bar{R}$ 
can give negative values, if the radius is not small enough,
because the vacuum expectation value of the Hermitian operator
$\underline{\hat{\bf E}}{^{\rm loc}}({\bf r},\omega)
\underline{\hat{\bf E}}{^{\rm loc\dagger}}({\bf r}_A,\omega)$ is not
necessarily positive. Figure~\ref{rmax} presents
$r_{\rm min}$ as a function of the damping parameter $\gamma$.
The curve was obtained numerically by requiring that 
$\Gamma^\perp$ must not be negative over the whole frequency spectrum.

Figure \ref{gamma001} shows that for chosen (small) $\gamma$ and $r$
$\!<$ $\!r_{\rm min}$ negative values of $\Gamma^\perp$ may appear 
when the atomic transition frequency 
$\omega_A$ approaches the medium resonance frequency $\omega_T$ and 
is in an interval that corresponds to the polariton bandgap between 
$\omega_T$ and $\omega_L$ $\!=$ $\![\omega_T^2 +(0.46 \omega_T)^2 ]^{1/2}$
in the Hopfield model of a dielectric in the absence of 
absorption \cite{hop58}. 
Obviously, in this regime of spontaneous decay, in which the decay rate 
is expected to respond to a change in the local-field noise very 
sensitively, the Clausius-Mosotti model has to be replaced with a
refined model, at least in quantum theory. 

From the standard derivation of the (classical) Clau\-sius-Mosotti
local-field (see, e.g., \cite{jackson}) the radius $\bar{R}$ of the
virtual cavity should be larger than the average distance of two
neighboring atoms but sufficiently smaller than the optical
wavelength. In terms of the parameter $r$, the latter requirement
means that $r$ $\!\gg$ $\!1$.
Provided that the damping parameter $\gamma$ is not 
too small, this is in agreement with the condition that the parameter
$r$ should not be smaller than  $\!r_{\rm min}$ given in Fig.~\ref{rmax}.

\section{Conclusions} \label{conclusions}
Within the frame of the Clausius-Mosotti model we have studied the 
influence of the quantum local-field correction arising from the noise 
polarization on the spontaneous decay rate of an excited atom embedded 
in an absorbing medium.
We have shown that inclusion in the local field of the 
noise polarization ensures that the local field fulfills
the fundamental equal-time commutation relations of QED,
provided that the static refractive index of the medium
does not exceed unity substantially.  
The calculated rates demonstrate that the contribution of the noise 
polarization to the local field is extremely important and cannot 
be ignored. In particular, at the resonance frequencies of the medium
the transverse decay rate can drastically change compared with the
classically corrected rate where the fluctuating component of the
polarization is omitted.  

The decay rate crucially depends on the choice of the
radius of the virtual cavity. It is worth noting
that from the dependence on the radius of the transverse rate -- an
effect that only results from the noise polarization -- a 
second condition of validity can be imposed on the local-field model.
In order to obtain for any transition frequency a positive  
transverse decay rate, the cavity radius must not exceed some 
upper bound.

The Clausius-Mosotti virtual-cavity model is commonly based on 
the assumption that the near field that arises from the atoms inside 
the cavity averages to zero in general. In quantum optics this assumption 
may fail, because of the vacuum noise associated with this field, which 
may be an explanation for the restrictions found. In order to extend the 
range of validity of the theory, a more refined concept seems to be 
necessary that considers the near field more carefully.  
\onecolumn
\appendix
\section{Commutation relations of the local field operators}
\label{appendix}
From Eq.~(\ref{22}) together with Eq.~(\ref{21}), the local 
electric field operator reads in Fourier space
\begin{equation}
\label{A1}
\underline{\hat{\bf E}}{^{\rm loc}}({\bf r},\omega) =
\frac{\epsilon({\bf r},\omega)+2}{3}\, \underline{\hat{\bf E}}({\bf
r},\omega) +\frac{i}{3\epsilon_0} \sqrt{\frac{\hbar\epsilon_0}{\pi}
\epsilon_I({\bf r},\omega) } \, \hat{\bf f}({\bf r},\omega) .
\end{equation}
Combining Eqs.~(\ref{24}) and (\ref{A1}), we obtain for the 
local magnetic field in Fourier space
\begin{equation}
\label{A2}
\underline{\hat{\bf B}}{^{\rm loc}}({\bf r'},\omega') = {\bf \nabla}
\times \Bigg[ {\cal P} \frac{1}{i\omega'} \, \underline{\hat{\bf
E}}({\bf r'},\omega') \, \frac{\epsilon({\bf r'},\omega')+2}{3}
+\frac{1}{3\epsilon_0} \,{\cal P} \frac{1}{\omega'}
\,\sqrt{\frac{\hbar\epsilon_0}{\pi} \epsilon_I({\bf r'},\omega')}
\,\hat{\bf f}({\bf r'},\omega') \Bigg] . 
\end{equation}
Recalling Eqs.~(\ref{25}) and (\ref{26}), from Eqs.~(\ref{A1})
and (\ref{A2}) together with Eq.~(\ref{15}) the local electric and 
magnetic fields are given by
\begin{eqnarray}
\label{A3}
\hat{E}_i{^{\!\!\rm loc}}({\bf r}) &=& \sqrt{\frac{\hbar\epsilon_0}{\pi}}
\int_0^\infty d\omega \int d^3{\bf s} \bigg[
\frac{\epsilon({\bf r},\omega)\!+\!2}{3\epsilon_0} \,\frac{i\omega^2}{c^2}
\,\sqrt{\epsilon_I({\bf s},\omega)} \,G_{ij}({\bf r},{\bf s},\omega)
\hat{f}_j({\bf s},\omega) +\mbox{H.c.} \bigg] 
\nonumber \\ && \hspace{0ex} 
+\,\frac{1}{3\epsilon_0} \,\sqrt{\frac{\hbar\epsilon_0}{\pi}}
\int_0^\infty d\omega \left[ i\sqrt{\epsilon_I({\bf r},\omega)}
\,\hat{f}_i({\bf r},\omega ) + \mbox{H.c.} \right] , 
\\
\label{A4}
\hat{B}_k{^{\!\!\rm loc}}({\bf r'}) &=& \epsilon_{klm} \partial_l^{r'}
\Bigg\{ \sqrt{\frac{\hbar\epsilon_0}{\pi}} \,{\cal P}\! \int_0^\infty
d\omega' \int d^3{\bf s'} \bigg[ \frac{\epsilon({\bf
r'},\omega')\!+\!2}{3\epsilon_0} 
\,\frac{\omega'}{c^2} \,\sqrt{\epsilon_I({\bf s'},\omega')} \,G_{mn}({\bf
r'},{\bf s'},\omega') \hat{f}_n({\bf s'},\omega') +\mbox{H.c.} \bigg]
\nonumber \\ && \hspace{0ex}
+\,\frac{1}{3\epsilon_0} \,\sqrt{\frac{\hbar\epsilon_0}{\pi}}
\,{\cal P}\!\int_0^\infty \frac{d\omega'}{\omega'} \left[ 
\sqrt{\epsilon_I({\bf r'},\omega')} \,\hat{f}_m({\bf r'},\omega') +
\mbox{H.c.} \right] \Bigg\} .
\end{eqnarray}
Thus, the (equal-time) commutator between the local electric and
magnetic fields can be given by
\begin{eqnarray} \label{A5}
\lefteqn{
\left[ \hat{E}_i^{\rm loc}({\bf r}), \hat{B}_k^{\rm loc}({\bf r'}) \right]}
\nonumber \\ &&\hspace{2ex}
=\,\frac{\hbar\epsilon_0}{\pi} \epsilon_{klm} \,\partial_l^{r'} \Bigg\{
{\cal P}\!\int_0^\infty \!d\omega \int d^3{\bf s} \bigg[
\frac{i\omega^3}{c^4}  
\,\frac{\epsilon({\bf r},\omega)\!+\!2}{3\epsilon_0}
\,\frac{\epsilon^\ast({\bf r'},\omega)\!+\!2}{3\epsilon_0}
\,\epsilon_I({\bf s},\omega) G_{ij}({\bf r},{\bf s},\omega)
G_{mj}^\ast({\bf r'},{\bf s},\omega) -\mbox{c.c.} \bigg] 
\nonumber \\&& \hspace{2ex}
+\,\frac{1}{3\epsilon_0} \,{\cal P}\!\int_0^\infty d\omega \bigg[
\frac{\epsilon^\ast({\bf r'},\omega)\!+\!2}{3\epsilon_0} \,\frac{i\omega}{c^2}
\,\epsilon_I({\bf r},\omega) G_{mi}^\ast({\bf r'},{\bf r},\omega)
-\mbox{c.c.} \bigg] 
\nonumber \\ && \hspace{2ex}
+\,\frac{1}{3\epsilon_0}\, {\cal P}\!\int_0^\infty d\omega \bigg[
\frac{\epsilon({\bf r},\omega)\!+\!2}{3\epsilon_0} \,\frac{i\omega}{c^2}
\,\epsilon_I({\bf r}',\omega) G_{im}({\bf r},{\bf r'},\omega)
-\mbox{c.c.} \bigg] 
\nonumber \\ &&\hspace{2ex}
+\,\frac{2i}{9\epsilon_0^2} \,{\cal P} \!\int_0^\infty
\frac{d\omega}{\omega} \left[ \epsilon_I({\bf r},\omega) \delta_{im}
\delta({\bf r}-{\bf r'}) \right] \Bigg\} .
\end{eqnarray}
The remaining spatial integral in Eq.~(\ref{A5}) can be calculated using the
symmetry relation
\begin{equation}
\label{A6}
G_{ij}({\bf r},{\bf r'},\omega) = G_{ji}({\bf r'},{\bf r},\omega),
\end{equation}
the crossing relation 
\begin{equation}
\label{A7}
G_{ij}({\bf r},{\bf r'},\omega)=G_{ij}^\ast({\bf r},{\bf r'},-\omega), 
\end{equation}
and the integral relation \cite{hoknowel98} 
\begin{equation}
\label{A8}
\frac{\omega^2}{c^2} \int d^3{\bf s} \, \epsilon_I({\bf s},\omega) 
G_{li}({\bf s},{\bf r},\omega) G_{lj}^\ast({\bf s},{\bf r'},\omega) 
= \frac{1}{2i} \left[ G_{ji}({\bf r'},{\bf r},\omega)
- G_{ij}^\ast({\bf r},{\bf r'},\omega) \right] .
\end{equation}
Straightforward calculation yields
\begin{eqnarray}
\label{A9}
\left[ \hat{E}_i^{\rm loc}({\bf r}), \hat{B}_k^{\rm loc}({\bf r'})\right]
= \frac{\hbar}{\pi\epsilon_0}\, \epsilon_{klm} \partial_l^{r'} 
\left[I^{(1)}_{im}({\bf r},{\bf r'}) + I^{(2)}_{im}({\bf r},{\bf r'}) \right],
\end{eqnarray}
where
\begin{eqnarray}
\label{A10}
I^{(1)}_{im}({\bf r},{\bf r'})
= {\cal P}\!\int_{-\infty}^\infty d\omega \,\frac{\omega}{c^2}\,
G_{im}({\bf r},{\bf r'},\omega) \left\{ 1+\textstyle\frac{1}{3} \left[
\epsilon({\bf r},\omega)\!-\!1 \right] +\frac{1}{3} 
\left[ \epsilon({\bf r'},\omega) \!-\!1 \right] 
+{\textstyle \frac{1}{9}} \left[ \epsilon({\bf r},\omega)\!-\!1 \right] 
\left[ \epsilon({\bf r'},\omega)-1 \right] \right\},
\end{eqnarray}
\begin{eqnarray}
\label{A11}
I^{(2)}_{im}({\bf r},{\bf r'})
={\cal P} \! \int_{-\infty}^\infty d\omega \,\frac{\epsilon_I({\bf
r},\omega)}{\omega} \,\delta_{im} \delta({\bf r}-{\bf r'}) .
\end{eqnarray}
Closing the integration contour in the upper complex frequency
half-plane and following the line in \cite{schknowel98}, we derive that
\begin{equation}
\label{A12}
I^{(1)}_{im}({\bf r},{\bf r'}) 
= i \pi \delta_{im} \delta({\bf r}-{\bf r'}) .
\end{equation}
Recalling the Kramers--Kronig relations, the $\omega$ integral
in Eq.~(\ref{A11}) is easily performed to obtain
\begin{equation}
\label{A13}
I^{(2)}_{im}({\bf r},{\bf r'})
=  \pi \left[ \epsilon_R({\bf r},0)\!-\!1 \right]
\delta_{im} \delta({\bf r}-{\bf r'})
= \pi \left[ \epsilon({\bf r},0)\!-\!1 \right]
\delta_{im} \delta({\bf r}-{\bf r'}).
\end{equation}
Combining Eqs.~(\ref{A9}), (\ref{A12}), and (\ref{A13})
then yields the commutation relation (\ref{27}).
\twocolumn



\begin{thebibliography}{99}

\bibitem{milonni}
P.W. Milonni, {\it The Quantum Vacuum: An Introduction to Quantum
Electrodynamics}, (Academic, San Diego, 1994).

\bibitem{glalew91}
R.J. Glauber and M. Lewenstein, Phys. Rev. A {\bf 43}, 467 (1991).

\bibitem{barhutlou92}
S.M. Barnett, B. Huttner, and R. Loudon, Phys. Rev. Lett. {\bf 68},
3698 (1992). 

\bibitem{barhutloumat96}
S.M. Barnett, B. Huttner, R. Loudon, and R. Matloob, J. Phys. B:
At. Mol. Opt. Phys. {\bf 29}, 3763 (1996).

\bibitem{juz97}
G. Juzeli\={u}nas, Phys. Rev A {\bf 55}, R4015 (1997).

\bibitem{tip97}
A. Tip, Phys. Rev. A {\bf 56}, 5022 (1997).

\bibitem{knomuk89}
J. Knoester and S. Mukamel, Phys. Rev. A {\bf 40}, 7065 (1989).

\bibitem{hulhilkle85}
R.G. Hulet, E.S. Hilfer, and D. Kleppner, Phys. Rev. Lett. {\bf 55},
2137 (1985).

\bibitem{jheandhinmesmoihar87}
W. Jhe, A. Anderson, E.A. Hinds, D. Meschede, L. Moi, and S. Haroche,
Phys. Rev. Lett. {\bf 58}, 1320 (1987).

\bibitem{kosshi96}
K. Koshino and A. Shimizu, Phys. Rev. A {\bf 53}, 4468 (1996).

\bibitem{takuji97}
I. Takahashi and K. Ujihara, Phys. Rev. A {\bf 56}, 2299 (1997).

\bibitem{tomlen97}
M.S. Toma\v{s} and Z. Lenac, Phys. Rev. A {\bf 56}, 4197 (1997).

\bibitem{yeugus96}
M.S. Yeung and T.K. Gustafson, Phys. Rev. A {\bf 54}, 5227 (1996).

\bibitem{barnes98}
W. Barnes, J. Mod. Opt. {\bf 45}, 661 (1998) and references cited
therein.

\bibitem{rikkes95}
G.L.J.A. Rikken and Y.A.R.R. Kessener, Phys. Rev. Lett. {\bf 74}, 880 (1995).

\bibitem{holland98}
F.J.P. Schuurmans, D.T.N. de Lang, G.H. Wegdam, R. Sprik, and
A. Lagendijk, Phys. Rev. Lett. {\bf 80}, 5077 (1998).

\bibitem{vrilag98}
P. de Vries and A. Lagendijk, Phys. Rev. Lett. {\bf 81}, 1381 (1998).

\bibitem{schknowel98}
S. Scheel, L. Kn\"oll, and D.-G. Welsch, Phys. Rev A {\bf 58}, 700
(1998).

\bibitem{gruwel96}
T. Gruner and D.-G. Welsch, Phys. Rev. A {\bf 53}, 1818 (1996).

\bibitem{hoknowel98}
Ho Trung Dung, L. Kn\"oll, and D.-G. Welsch, Phys. Rev. A {\bf 57},
3931 (1998). 

\bibitem{matloubarjef95}
R. Matloob, R. Loudon, S.M. Barnett, and J. Jeffers, Phys. Rev. A {\bf
52}, 4823 (1995). R. Matloob and R. Loudon, Phys. Rev. A {\bf 53},
4567 (1996).

\bibitem{hutbar92}
B. Huttner and S.M. Barnett, Phys. Rev. A {\bf 46}, 4306 (1992).

\bibitem{hop58}
J.J. Hopfield, Phys. Rev. {\bf 112}, 1555 (1958).

\bibitem{fan56}
U. Fano, Phys. Rev. A {\bf 103}, 1202 (1956).

\bibitem{jackson}
J.D. Jackson, {\it Classical Electrodynamics}, (Wiley, New York, 1975).

\bibitem{bowdow93}
C.M. Bowden and J.P. Dowling, Phys. Rev. A {\bf 47}, 1247 (1993).

\bibitem{yab87}
E. Yablonovitch, Phys. Rev Lett. {\bf 58}, 2059 (1987).

\end{thebibliography}
\end{document}